# A Novel Multivariate Model Based on Dominant Factor for Laser-induced Breakdown Spectroscopy Measurements

Zhe Wang, Jie Feng, Lizhi Li, Weidou Ni, Zheng Li[*]

*State Key Lab of Power Systems, Department of Thermal Engineering, Tsinghua-BP Clean Energy Center, Tsinghua University, Beijing, China*

**Abstract:**

This paper presents a new approach of applying partial least squares method combined with a physical principle based dominant factor. The characteristic line intensity of the specific element was taken to build up the dominant factor to reflect the major elemental concentration and partial least squares (PLS) approach was then applied to further improve the model accuracy. The deviation evolution of characteristic line intensity from the ideal condition was depicted and according to the deviation understanding, efforts were taken to model the non-linear self-absorption and inter-element interference effects to improve the accuracy of dominant factor model. With a dominant factor to carry the main quantitative information, the novel multivariate model combines advantages of both the conventional univariate and PLS models and partially avoids the overuse of the unrelated noise in the spectrum for PLS application. The dominant factor makes the combination model more robust over a wide concentration range and PLS application improves the model accuracy for samples with matrices within the calibration sample set. Results show that RMSEP of the final dominant factor based PLS model decreased to 2.33% from 5.25% when using the conventional PLS approach with full spectral information. Furthermore, with the development in understanding the physics of the laser-induced plasma, there is potential to easily improve the accuracy of the dominant factor model as well as the proposed novel multivariate model.

## 1 Introduction

Laser-induced breakdown spectroscopy (LIBS) is an atomic emission technique by focusing a high-power, short-pulse laser on the sample surface to form a plasma and analyzing the emitted spectrum to obtain the chemical composition of the sample. LIBS technology exhibits numerous appealing features that distinguish it from conventional analytical spectrochemical techniques, such as little to no sample preparation and capability to analyze any kind of sample. It has been applied to remote sensing [1-3], forensic analysis [4], ceramic raw materials [5], wood products [6], determination of major or minor elements in metal [7-8], coal analysis [9-10] and many other fields. Commercial analytical instrument has also been developed to provide LIBS measurement on laboratory bench [11-12].

Currently, there are two methods, conventional univariate method and PLS method, applied for LIBS quantitative measurement generally. The conventional univariate model of LIBS directly connects the intensity of the specific element with its concentration based on the fact that the more species in the plasma, the higher the measured characteristic line intensity. The model can generally be applied over a wide concentration range because of its physical background. The model is also the most commonly applied model currently due to its simplicity. However, precise quantitative analysis of LIBS is very complicated due to uncontrollable fluctuations of the experimental parameters and the physical and

---

[*] Corresponding author. Tel: +86 10 62795739; fax: +86 10 62795736

E-mail address: lz-dte@mails.tsinghua.edu.cn

chemical matrix effects [13]. Various fluctuations will weaken the theoretical relationship between the intensity and the elemental concentration [14], deteriorating the measurement accuracy.

A new and promising method for LIBS data interpretation is to utilize multivariate analysis to extract more quantitative information from the entire spectrum or a bunch of spectral lines of the sample instead of only one specific line intensity as in univariate model. PLS is such an approach and has shown great potential in recent years for LIBS measurements [15-21]. Generally, the PLS model has higher accuracy in predicting the elemental concentration. However, the PLS method was constructed based on statistical correlation between the measured spectra and the set of samples for calibration, while the physics principles are almost neglected. Therefore, the prediction of PLS model is not so accurate if the nature of matrix of the measured samples varies from the calibration sample set [13]. As shown by Fink et al. [22], the relative prediction error for all elements (Ti, Sb, and Sn) in recycled thermoplasts applying PLS method was typically in the order of 15-25% due to differences in the matrix and inconstant ablation behaviour. Another limitation inherent to the linear nature of PLS is that it could not satisfactorily model the non-linear relationship between the spectrum and the species concentration such as the saturation effects of the signal due to self-absorption of strong lines in the plasma. Sirven et al. [23] found that PLS was outweighed by artificial neural network (ANN) due to the linear modeling nature of PLS. Generally, the common way of applying PLS is to input the whole spectrum for the calibration [17]. The excess of information present in the spectrum or signal of noise, most of which are unrelated to the elemental composition, might worsen the calibration model, because the redundancy may add more uncertainties to the parameters calculated by PLS and therefore deteriorate the model robustness [24-26].

This paper presents a new approach of applying PLS method based on the understanding of the physical principles of plasma and the linear nature of the PLS approach. The major part of the concentration is extracted from the characteristic line intensity of the specific element explicitly as the dominant factor, while PLS approach is further applied to minimize the residual errors by utilizing more spectral information to compensate for the fluctuations of plasma. In essence, this dominant-factor based multivariate model combines advantages of both the univariate and PLS models. By utilizing a dominant factor to contain the main quantitative information, the model avoids the overuse of the unrelated noise in the spectrum and becomes more robust over a wide concentration range. The application of PLS to the residual error helps to improve the model accuracy within the matrix of calibration sample set.

## 2 Model descriptions

For LIBS measurements, the laser-induced plasma is typically at a state known as local thermodynamic equilibrium (LTE), in which, along with the detectors typically used in LIBS measurements, the integrated characteristic line intensity is given by [1]:

$$I_{ij} = F n_i^s A_{ij} = F \frac{A_{ij} g_i}{U^s(T)} n^s e^{-E_i/kT} \qquad (1)$$

where $F$ is an experimental parameter that takes into account the optical efficiency of the collection system as well as the plasma density and volume, $n_i^s$ indicates the number density of the species $s$ at excited level $i$, $A_{ij}$ is the transition probability, $g_i$ and $E_i$ are the statistical weight and the excitation energy for the excited level, respectively, $n^s$ is the total number density of the element $s$, $k$ is the Boltzmann constant, and $U^s(T)$ is the internal partition function of element $s$ at temperature $T$, which can be derived from NIST [27].

Under the conditions of stoichiometric ablation and constant plasma property, which can be characterized using plasma temperature, electron density and total elemental number density [28-29], **Eq.**1 can be simplified as:

$$I_{ij} = K C_s \qquad (2)$$

where $C_s$ is the elemental concentration and $K$ is a constant. This means that ideally, the measured line intensity is proportional to the species concentration in the sample as shown as the dash line with hollow squares in **Fig.**1. Conventional univariate model, $I_{ij} = KC_s + d$, is therefore built up based on this understanding by using the constant $d$ to describe the existing drift.

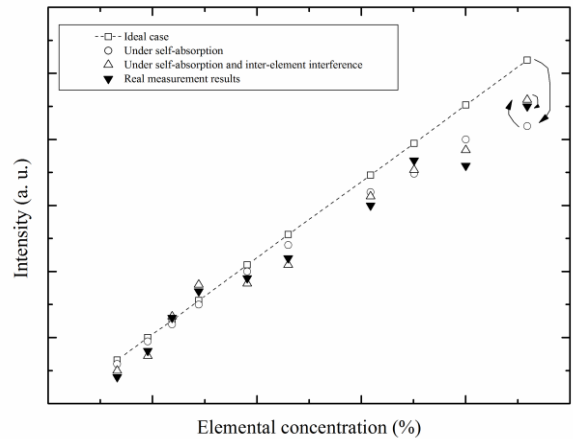

**Figure 1. The evolution of the characteristic intensity along with the elemental concentration change in LIBS measurement**

However, the intensity deviates from the ideal value due to different influencing factors and processes. Self-absorption is often unavoidable in LIBS quantitative analysis if the concentration of measured species is not low enough. Atoms at the lower energy levels can easily reabsorb the radiation emitted by other atoms of the same species in the plasma. This leads to a pronounced non-linear relationship between the line intensity and the increasing element concentration [30]. That is, the characteristic line intensity will differ from the ideal straight line as the element concentration increases. As shown in **Fig.** 1, the hollow circles show that as the elemental concentration increases, the increase of the line intensity slows down. Accordingly, this indicates that $K$ in **Eq.** 2 is no longer a constant and it changes with the elemental concentration.

Inter-element interference due to line overlap and the matrix effect is also unavoidable, especially for multi-element samples. Spectral interference is prominent when emission lines of other elements are close to an emission line of the analyte. In such situations, the characteristic line intensity might not only result from the transition of one single species, but also be interrupted by other elemental number densities. The effects of interference are of complexity, and the detected intensity may further deviate from the idea straight line, as shown as the hollow triangles in **Fig.** 1.

Besides, many other factors and processes would also alter the measured characteristic line intensity. The factors such as the power of laser, lens-to-sample surface distance and delay time, fluctuate from pulse to pulse, leading to the fluctuations of plasma itself. Although the fluctuations can generally be minimized by averaging the measured signal for multi-pulse, the deviation of the measured line intensity from expected value is still unavoidable. Furthermore, there may be other deviations due to the fluctuation of spatial and temporal transient transformation process of the laser-induced plasma. All these effects further depart the measured line intensity from the hollow triangle points to the dark triangles, which stand for the real measured value, as shown as in **Fig.** 1. It should be noticed that all these deviation processes shift the line intensity simultaneously, making it very difficult to separate them one by one physically and indicating that the utilization of data processing technology to compensate for these effects accordingly can be an effective way to improve the measurement results.

Due to these deviations, the intensity of the characteristic lines may not carry enough information to accurately reflect the measured element concentration, while it still contains the most correlated information. Therefore, an ideal way is to extract the major concentration information from these characteristic lines and further correct the model by taking the full-range spectrum into account to compensate for the deviations. In the present work, a PLS model based on the dominant factor was proposed with the understanding of the deviation evolution, in which the major part of the element concentration was extracted explicitly from characteristic line intensity of the measured element, or optionally, some characteristic line intensities of another element in the sample. The explicitly extracted expression for element concentration calculation is called "dominant factor" since it takes a dominant portion of the total model results. The details of the approach are explained as follows.

Since PLS method can tackle the linear relation between the line intensities and the elemental concentrations, to extract the dominant factor using the ideal linear equation (**Eq.** 2) may be not necessary and the non-linear self-absorption model is preferred presently. In the present work, the following empirical expression was applied [31]:

$$C_i = C_0 \ln(\frac{bC_0}{a+bC_0-I_i}) \qquad (3)$$

where $I_i$ is the characteristic line intensity; $b$ and $a$ are the constants calculated by the curve fitting technology, $C_0$ can be regarded as "saturation concentration", and $C_i$ is the predicted elemental concentration. By applying this self-absorption model, the results now correspond to the hollow circle points as shown in **Fig.** 1.

The mechanism of inter-element interference is very complicated and still remains unclear. There has been very little work to model the phenomenon in the literatures. In the present work, the residual error, which is defined as the difference between the model predicted concentration and the real elemental concentration in the sample and under current situation, the difference between the self-absorption model result and the real elemental concentration, was further minimized by modeling the inter-element interference effect using curve fitting technology [32]. It needs to be pointed out that non-linear correction was preferred to model the interference effect in our approach since it is believed that PLS method can tackle the linear relations effectively. After this process, the model results now correspond to the hollow triangle points as shown in **Fig.**1.

After the inter-element interference correction, the remaining deviation may mainly come from the imperfectness of the physical model in describing the relation between line intensity and elemental concentration, different fluctuations and other unknown factors, making it difficult to explicitly model these effects. A logical way is to utilize the full spectrum to further minimize the deviation. Currently popularly applied multivariate PLS approach is such a good candidate. Basically, PLS is a technique for modeling a linear relationship between a set of output variables and a set of input variables. Firstly, PLS creates uncorrelated latent variables which are linear combinations of the original input variables. A least squares regression is then performed on the subset of extracted latent variables [33]. In conventional LIBS application, PLS generates a regression model that correlates the two matrices, the LIBS spectra ($X$) and the elemental concentrations ($Y$) as described by **Eq.** 4.

$$Y = XB \qquad (4)$$

where $Y$ contains the elemental concentrations for each element (the response) in each calibration sample and $X$ includes the intensity of each wavelength for each calibration sample. $B$ is the regression coefficient matrix. As a result, the PLS analysis obtains a linear combination of values to correlate the spectral intensities with the elemental composition as follows [13]:

$$Y = b_0 + b_1 x_1 + ... + b_k x_k \qquad (5)$$

where $Y$ is the elemental concentration, $x_n$ is the spectral intensities at different wavelength, $b_n$ is the regression coefficient. As seen above, since the full spectral information has been utilized in PLS approach, that is, there is more variables for the calibration and prediction in PLS; therefore, it is more flexible to compensate for the fluctuations for PLS. Generally, PLS has advantage in calibration and prediction over univariate method.

In the present application, the PLS method was applied to model the residual error instead of the generally applied total elemental concentration with the full spectral range. That is, if we have extracted the dominant factor by considering the self-absorption and interference effects, the PLS method will be only applied to compensate for the differences from the hollow triangle points to the dark triangles as shown in Fig.1. Basically, the present dominant-factor PLS model is established mostly based upon the physical principles while still keeping the advantages of the multivariate PLS approach. Compared with the general PLS model, the new PLS model should be more applicable for a wider matrix range due to its physical background. Moreover, presently PLS is only applied to compensate for the relatively much smaller deviation compared with its generally application mode for the whole elemental concentration. That is, in general PLS method, linear relation is directly used to fit non-linear spectra-to-concentration curve; while here linear PLS is only applied to fit the much smaller non-linear spectra-to-residuals curve. This should lead to a better fit and better model results. In addition, this is further improved by explicitly extracting part of the non-linear relation between the spectra and the concentration in the dominant factor calculations.

## 3 Experimental setup

The instrument used for the present study was the Spectrolaser 4000 (XRF, Australia). More details and the schematic representation about the instrument are showed in our previous paper [29]. The detection system was composed of 4 Czerny-Turner spectrometers and CCD detectors which cover the spectral range from 190 to 940 nm with a nominal resolution of 0.09 nm. The broadband spectral response means that LIBS is capable of detecting all chemical elements, since all elements emit light somewhere in that spectral range [23]. The sample was placed on an auto-controlled X-Y translation stage.

Standardized brass samples from Central Iron and Steel Research Institute (CISRI) of China were chosen for the experiment, since they are highly homogenous and calibrated accurately. **Table** 1 shows the elemental concentrations of 14 standard brass alloys from CISRI of China used in the experiment.

**Table 1 The major elemental concentration of the samples**

| Samples NO. | Cu (%) | Pb (%) | Zn (%) | Fe (%) | P (%) | Sn (%) | Sb(%) |
|---|---|---|---|---|---|---|---|
| **ZBY901** | 73.00 | 2.77 | 23.99 | 0.028 | 0.0043 | 0.019 | 0.0036 |
| **ZBY902** | 64.43 | 1.87 | 33.45 | 0.036 | 0.012 | 0.032 | 0.0034 |
| **ZBY903** | 60.28 | 0.766 | 38.79 | 0.047 | 0.0042 | 0.108 | 0.0061 |
| **ZBY904** | 59.14 | 1.50 | 38.85 | 0.167 | 0.011 | 0.102 | 0.0077 |
| **ZBY905** | 58.07 | 1.81 | 39.59 | 0.110 | 0.020 | 0.269 | 0.013 |
| **ZBY906** | 56.62 | 0.581 | 41.76 | 0.037 | 0.044 | 0.478 | 0.022 |
| **ZBY907** | 59.55 | 3.06 | 34.92 | 0.502 | 0.020 | 0.750 | 0.029 |
| **ZBY921** | 59.89 | 0.318 | 39.01 | 0.288 | 0.084 | 0.0046 | 0.023 |
| **ZBY922** | 61.88 | 0.108 | 37.53 | 0.116 | 0.039 | 0.0051 | 0.0046 |
| **ZBY923** | 69.08 | 0.018 | 30.44 | 0.052 | 0.011 | 0.0081 | 0.0072 |
| **ZBY924** | 80.90 | 0.017 | 18.75 | 0.110 | 0.013 | 0.010 | 0.010 |
| **ZBY925** | 85.06 | 0.029 | 14.79 | 0.028 | 0.0052 | 0.011 | 0.0091 |
| **ZBY926** | 90.02 | 0.0084 | 9.76 | 0.024 | 0.0071 | <0.0010 | 0.0031 |
| **ZBY927** | 95.90 | 0.0028 | 4.02 | 0.012 | 0.0046 | <0.0010 | 0.0013 |

Before the analysis, the sample surface was carefully cleaned using ethanol and then dried in air to remove contaminations. The samples were placed in the sample chamber and exposed to air with pressure very close to 1 atm. The machine warm-up time was set to be more than half hour since it was found that signal fluctuations can be greatly reduced with longer warm-up time. The laser energy and delay time were optimized to be 90 mJ/pulse and 2.25 μs, respectively. These parameters could produce spectra with negligible Bremsstrahlung radiation and without saturating line intensity to the spectrometers. A laser pulse with higher energy (150 mJ) was applied to burn off the contaminations. In order to average out the experimental parameter fluctuations and the sample heterogeneity, an averaged spectrum of 35 replications at different locations on the sample surface was obtained for each sample.

The pre-treatment of the spectra includes subtracting the background signal, which comes from the instrumental and environmental noise and was recorded by the spectroscopy with a long enough delay time and a laser pulse with much lower energy, to reduce the unintended measurement errors. Additionally, the spectra were corrected for the efficiency of the detection system to minimize the line intensity distortion from the wavelength dependant efficiency of collecting optics, lenses and fibre optics, the spectrometer gating, the detector sensor and intensifier. To further remove out the unexpected fluctuations, the spectra were normalized with the whole spectrum area.

## 4 Results and discussion

In the following sections, the new PLS model with different dominant factors were evaluated in terms of Cu concentration determination together with the conventional PLS model. To clearly demonstrate the improvement made by our approach, the conventional PLS with the full spectral range input were chosen as the baseline. The software SPSS 17.0 (SPSS Inc., Chicago, Illinois, USA) was used to perform the PLS calculation. The coefficient of determination ($R^2$) and the root mean square error of prediction (RMSEP) were used to assess the performance of the calibration and prediction qualities of the model. $R^2$ determines the quality of the calibration curve and RMSEP is a measurement of the average prediction errors for the validation set. An ideal model will have a $R^2$ close to 1 and RMSEP values close to 0. Moreover, to evaluate the overall performance of the models, root mean square error of calibration and prediction (RMSEC&P) is proposed in the present work, as a smaller RMSEC&P indicates a better model quality. Ten samples were selected to build the calibration model and four samples (ZBY906, ZBY907, ZBY924 and ZBY927) were used to evaluate the model prediction. These four samples were chosen purposely since Cu concentrations of ZBY907 and ZBY924 are within the concentration range of the calibration samples and Cu concentrations of ZBY906 and ZBY927 are out of the calibration sample set. It is believed that the new proposed PLS model is applicable to a wider sample matrix.

### 4.1 Baseline

PLS is one of the advanced analytical tools in the field of chemometrics and it has been proven to work effectively for situations where the number of observations is large and high multi-collinearity among the variables exists. LIBS is under such conditions and PLS has shown great potential in LIBS quantitative measurements. Normally, PLS has a better results than conventional univariate model, therefore, it was chosen as the baseline. In the baseline PLS model, the full spectral range and Cu concentrations of samples were used as input variables and the number of principle components was optimized to be 3.

Basically, the PLS method applies the full spectral information to implicitly compensate for the signal deviation due to fluctuations of temperature, electron number density, and total number density, self-absorption, inter-element interference using linear correlations, and normally it yields very good calibration results. Since PLS is not able to accurately reflect the non-linear relation between the spectral intensity and the sample elemental concentrations, as the matrix of the measured sample is out of the calibration sample set, the prediction may not be satisfactory. As shown in **Fig.** 2, the $R^2$ of the PLS model is 0.999, the RMSEP is as high as 5.25%, and the RMSEC&P is 2.81%. It was found that the prediction errors for ZBY907 and ZBY927 were the major sources for the RMSEP value. This partly resulted from the fact that the conventional PLS method neglects physical principles and confirms that the conventional PLS method requires the matrix of the measured samples within the calibration sample set since ZBY927 has the higher Cu concentration than all calibration sample while ZBY907 is at the lowest end of the Cu concentration of the calibration set. Another reason for ZBY907 to have large measurement error may come from the non-linear inter-elemental interference since ZBY907 has the highest Pb concentration among all samples as listed in **Table** 1 and Pb species in the plasma might show non-linear inter-element interference for Cu characteristic line intensity.

The results demonstrate that linear PLS model may not satisfactorily model the non-linear effects in LIBS quantitative measurement. Moreover, the contributions of noise are widely presented in the spectra. The full spectral range input might contain too much noise and make the PLS model less robust due to the excess of unrelated information, which is another disadvantage of conventional PLS. Establishing a PLS model with a dominant factor based on physical principles is a potential way to avoid overuse of the noise.

In addition, the conventional PLS treats every spectrum evenly without extracting a dominant factor based on the intensities of Cu characteristic lines. Since the characteristic line intensity contains much more information than other lines to the specific elements, it is reasonable to take more weight from the characteristic lines, which is the basic logic for the dominant factor PLS approach.

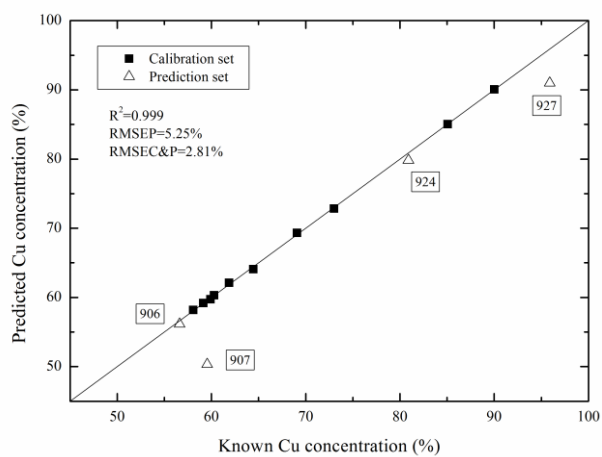

**Figure 2. Baseline model results**

**4.2 Dominant factor extraction**

The proposed novel PLS model is mainly determined by the dominant factor based on physical principles. Therefore, any development in the understanding of the physics of the LIBS helps to improve the dominant factor accuracy and then the model results. In the following section, different dominant factors were calculated mainly to explicitly handle the non-linear effect which cannot be simply compensated for by the linear PLS approach.

The simplest dominant factor is extracted only from the characteristic line intensity of the specific element using linear relation (conventional univariate model). The spectral area of Cu(I) line at 570.024 nm was used to construct the dominant factor since its raw data provided the best curve-fitting results. Figure 3 shows the calibration and prediction results of the univariate linear model (called dominant factor 1). Though the characteristic line intensity is the most sensitive variable to reflect the elemental concentration theoretically, $R^2$ of the linear model is only 0.913, showing that such a conventional linear dominant factor is often insufficient to provide accurate reflect. As shown in Fig. 3, RMSEP is 3.80%, which is better than the baseline PLS model, showing that a model based on physical principle may be more robust than PLS method under conditions where the matrix of the measured sample is out of the calibration set. Overall, the RMSEC&P of dominant factor 1 is 3.36%, which is much larger than the baseline PLS model. According to **Fig.** 1, the PLS method applied with the dominant factor 1 compensate for the difference between the real elemental concentration (the dark triangles) and the ideal cases (the hollow squares).

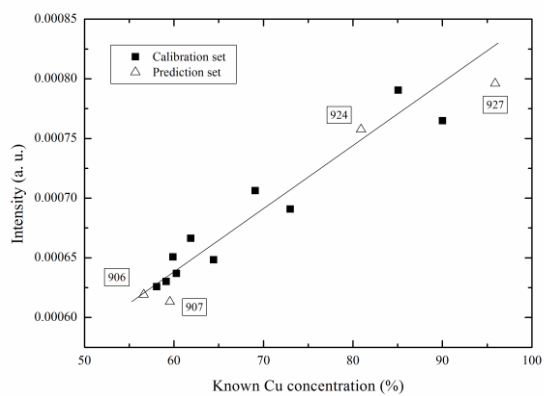

(a)

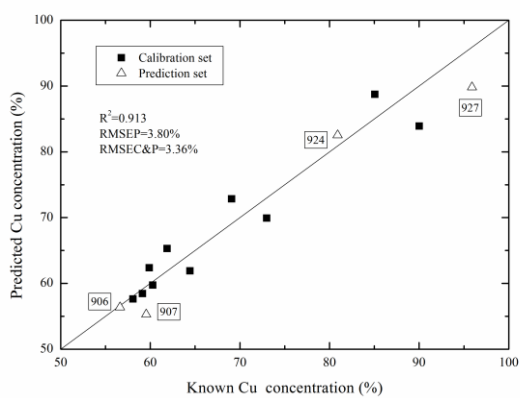

(b)

**Figure 3. Dominant factor 1 results: (a) characteristic line intensity versus Cu concentration; (b) calibration and prediction results**

As PLS only applies linear relations to correlate the spectral intensities and element concentrations, its combination with the linear dominant factor 1 is not able to reflect the non-linear effect. A reasonable way to handle the non-linear effects is using the dominant factor. In the following, the dominant factor taking the non-linear effect, mainly self-absorption and inter-element interference, were discussed.

The dominant factor considering the self-absorption effect was obtained using the characteristic line intensity at 570.024 nm of the interested element (Cu). That is, **Eq.** 3, $C_0 \ln(\frac{bC_0}{a+bC_0-I_i})$, is utilized to construct the dominant factor (called dominant factor 2). The parameter $C_0$ in **Eq.** 3 was estimated to be 0.5123 using best curve fitting technology. $R^2$ of the fitting was improved to 0.919, RMSEP was decreased to 2.39%, and RMSEC&P was reduced to 3.01%, as shown in **Fig.** 4. The absolute relative errors for ZBY927 decreases to 1.47% compared with 6.32% as in the linear model (dominant factor 1), but the absolute relative error of ZBY907 is still as high as 6.23%. According to **Fig.** 1, the deviations between the hollow triangle points and the hollow squares needs to be compensated for by PLS method with dominant factor 2. Comparing the results of dominant factor 1 and dominant factor 2, the deviations between the dominant factor model and the real values were much reduced, therefore, there are smaller deviations needed to be corrected by the linear PLS approach and there should be better results for the proposed multivariate with dominant factor.

According to the model description in **Fig.** 1, the left deviations should mainly from inter-element interference and other fluctuations. Therefore, after the consideration of self-absorption, the next step is to apply appropriate to describe inter-element interference and make the model more robust.

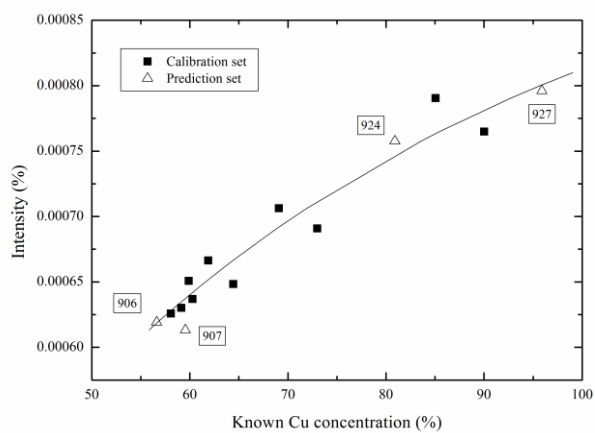

(a)

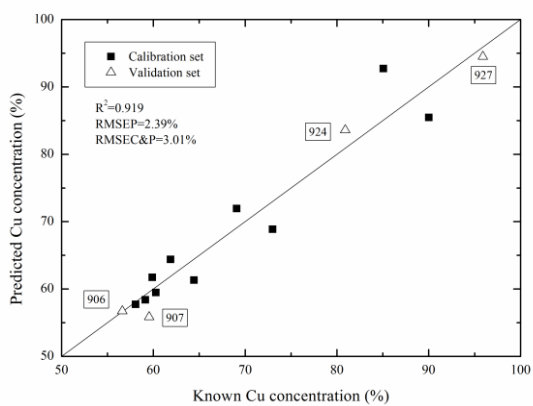

(b)

**Figure 4. Dominant factor 2 results: (a) characteristic line intensity versus Cu concentration; (b) calibration and prediction results**

The mechanism of inter-element interference is very complicated and there has not been a way to model the inference effects from physical mechanism. In present work, the inter-element interference effects to the Cu line intensity (570.024 nm) were correlated to other elements' line intensities. The correlation between the residual errors of the dominant factor 2 with various characteristic line intensities of other elements were tested and the peak area of Pb(I) intensity at 363.957 nm was found to have the best correlation, which means that Pb may be the main species in influencing the Cu line intensity at 570.024 nm. Therefore, the peak area of Pb(I) was taken to further improve the dominant factor. The residuals of dominant factor 2 and versus Pb(I) intensity at 363.957 nm was shown in **Fig.** 5, in which there is not a clear relation. This is mainly due to the fact that not only the species Pb(I) but also others affect the Cu characteristic line intensity and the line intensity of Pb(I) was influenced by species Cu and others. Moreover, the self-absorption model may exist some uncertainty. The purpose of the present work is not to develop a thorough model to describe the inference effect but to compensate partly for the interference effect using curve fitting technology.

Since any linear correlation between the characteristic of different species can be compensated for implicitly during the PLS residual errors correction, only non-linear relation were taken to improve the dominant factor. In the present work, a quadratic polynomial equation was obtained to further improve the dominant factor using the best curving fitting technology.

$$E_{Cu} = a_0 + a_1 I_{Pb} + a_2 I_{Pb}^2 \qquad (6)$$

where $E_{Cu}$ is the residual errors of dominant 2, and $a_0$, $a_1$, $a_2$ are constants obtained from curve fitting technology. Now the dominant factor (dominant factor 3) can be written as:

$$C_i = C_0 \ln(\frac{bC_0}{a + bC_0 - I_i}) + a_0 + a_1 I_{Pb} + a_2 I_{Pb}^2 \qquad (7)$$

In addition, it was found that if a cubic polynomial equation was applied to compensate for the inter-element interference, the final dominant factor based PLS methods yields even a better result. However, it may be not physically possible to have a cubic polynomial relation for the inter-element interference between the Pb(I) and Cu(I). Therefore, it was not chosen as preferred. Based on the scenarios listed in **Fig.** 1, after extracting the inter-element interference, only the deviations between the circle points and the hollow triangle points needs to be compensated by PLS approach.

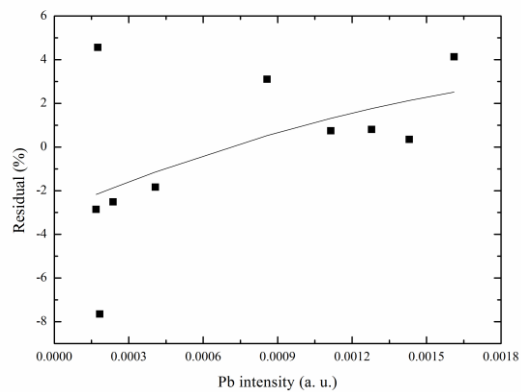

**Figure 5. Correlation between Cu residual errors and Pb intensity**

The calibration and prediction results of dominant factor 3 are shown in **Fig.** 6. All results are better than those of dominant factor 2. The absolute relative error for ZBY907, in which there is highest Pb concentration and might result in strong inter-element interference, was reduced to 1.04% from 6.23% as in dominant factor 2.

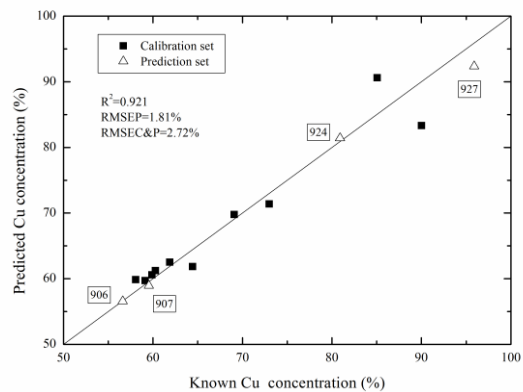

**Figure 6. Dominant factor 3 results**

Dominant factor modelled either linear or non-linear relation between the characteristic line intensities and the concentration of the specific element, which partially made up the limitation of the PLS approach such as linear correlation and non-physical background. However, the $R^2$ of the all dominant factor models are not close to unity enough because other fluctuations such as plasma properties have not yet been compensated for. Since there is correlation between these fluctuations and different emission lines, PLS approach can be used to implicitly further minimize the residual errors of the dominant factors. In addition, comparing the three dominant factors, it is confirmed that the more the physical processed is considered, the more robust the dominant factor model is. Furthermore, as smaller residual error needs to be corrected by PLS, the possibility of the overfitting was reduced and it helps to improve the model accuracy. As seen above, the three dominant factors normally have a better prediction results (RMSEP) compared with the baseline PLS model. This is mainly because the dominant factors were built up based on physical principle, making it more robust for situations where the matrix of the predicted sample were out of the calibration set. According to the value of the RMSEC&P, dominant factor 3 performs better than the baseline PLS model, which partly proves our method in modeling the inter-element interference effect. The results of different dominant factors were listed in **Table** 2.

Table 2. List of dominant factors

| Models | Dominant Factors | $R^2$ | RMSEP (%) | RMSEC&P (%) |
|---|---|---|---|---|
| Baseline | No | 0.999 | 5.25 | 2.81 |
| Dominant factor 1 | $C_i = AI_i + D$ (1) | 0.913 | 3.80 | 3.36 |
| Dominant factor 2 | $C_i = C_0 \ln(\frac{bC_0}{a+bC_0-I_i})$ (2) | 0.919 | 2.39 | 3.01 |
| Dominant factor 3 | $C_i = C_0 \ln(\frac{bC_0}{a+bC_0-I_i}) + g(I_{Pb})$ (3) | 0.921 | 1.81 | 2.72 |

(1) $A=1.889\times 10^3$, $D=-0.6063$
(2) $C_0=0.5123$, $a=-6.827\times 10^{-5}$, $b=2.004\times 10^{-3}$
(3) $g(I_{Pb}) = a_0 + a_1 I_{Pb} + a_2 I_{Pb}^2$, $a_0=-2.960\times 10^{-2}$, $a_1=48.22$, $a_2=-8.822\times 10^3$

It is necessary to mention that since the influence of the inter-element interference effect to the characteristic line intensity of the measured species is small compared with characteristic line intensity itself. Therefore, the interference effect can be very obscure using the raw spectral data because it is submerged within all the deviations which occur simultaneously and interlaced together with interference process. However, if the other effects can be explicitly separated, the inter-element interference can be much clearer and possibly be modelled as described above. The more accurately the other deviation effects were modelled, the clearer the inter-element interference can be found.

**4.3 PLS correction based on dominant factor model**

The residual errors of the dominant factor, which may be mainly from plasma properties fluctuations, can be further minimized with PLS approach as describe above. In the followings, the final results of PLS method based on different dominant factors were discussed.

With the dominant factor 1 and PLS approach to compensate for the residual, $R^2$ is 0.999 and RMSEP is 3.28% as shown in **Fig.** 7. Compared with the results of dominant factor 1 itself, $R^2$, representing the quality of the calibration curve has been improved greatly mainly due to the PLS residual correction, while RMSEP, representing the quality of the model prediction, is only improved from 3.80% to 3.28% since the dominant factor itself determines the major part of model results. For overall performance, the RMSEC&P was decreased to 1.76%, showing that the combination of the dominant factor with PLS correction significantly improve the final results. Compared with the baseline model, $R^2$ is the same while RMSEP decreases significantly. That is, by combining the advantage of the univariate model and PLS method, the proposed PLS model is not only comparable with PLS method for samples having matrix within the calibration set, but also have the same robustness in measuring samples out of the calibration set as the conventional univariate model based on physical principle. Therefore, the RMSEC&P was declined to 1.76% from 2.81% as in the baseline PLS model, demonstrating the proposed model obviously outweighs the conventional PLS.

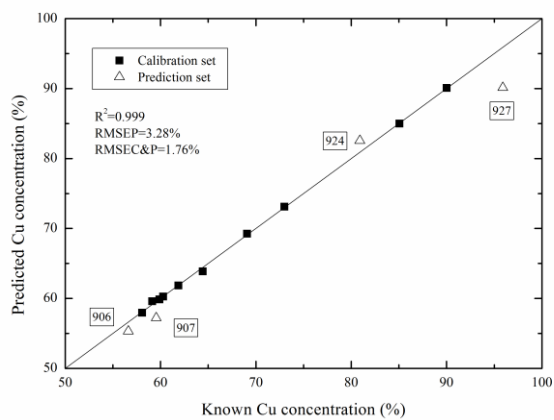

**Figure 7. Final results of residual correction PLS model with dominant factor 1**

Based on the non-linear dominant factor 2, the final results of the PLS residual correction model yields even better prediction results while keeping the same calibration quality compared with that of dominant factor 1. As shown in **Fig.** 8, the RMSEC&P declines to 1.43% and RMSEP decreases to 2.63%, which are both better than that of both the baseline model and the PLS residual error correction model with dominant factor 1. This is mainly because the non-linear self-absorption effect was explicitly extracted, making the sample ZBY927 with the highest Cu concentration among all samples less relative measurement error. Because in the process to correct the residual errors, PLS method tries to compensate for any deviations using linear correlation and full spectral information, there could be noise overfitting problem, making RMSEP of the final model (2.63%) larger than the dominant factor 2 itself (2.39%).

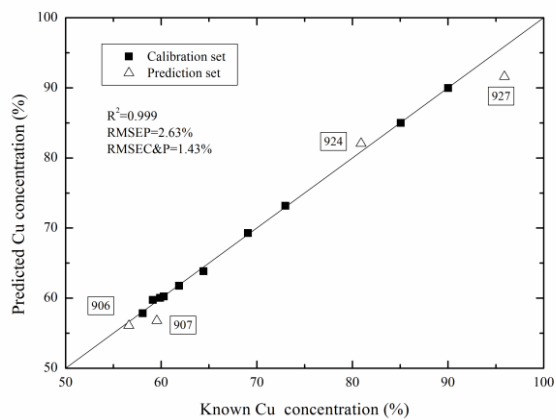

Furthermore, if inter-element interference is modeled as in dominant factor 3, the PLS correction model is more accurate as shown in **Fig.** 9. RMSEP is further reduced to 2.33%, showing better prediction accuracy than that of dominant factor 2, while the model calibration quality remains the same. For ZBY907 measurement, the result of dominant 3 is much better than that of dominant factor 2 since it further modelled the inter-element interference effect. However, the result of the combination approach with dominant factor 3 does not show much more accuracy over that of with dominant factor 2 in predicting the Cu concentration of ZBY907 as shown in **Figs.** 7 and 8. This might come from the intrinsic limitations of the PLS approach, including noise overfitting and applying the linear correlation to compensate for the non-linear relation between line intensity and plasma properties fluctuations.

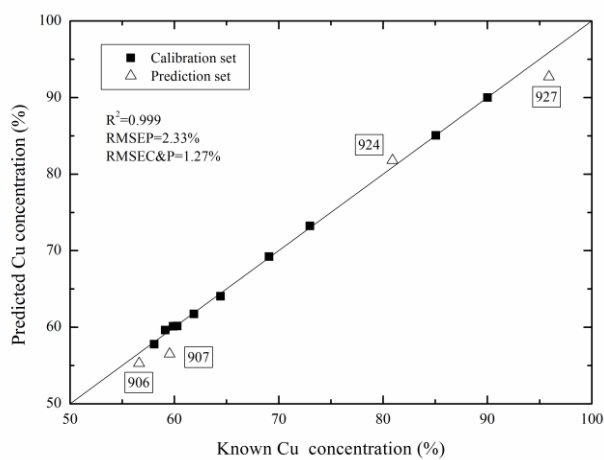

**Figure 9. Final results of residual correction PLS model with dominant factor 3**

The PLS models with different dominant factor are listed in **Table 3**, together with the baseline PLS model.

**Table 3. Results of PLS models with different dominant factors**

| Models | $R^2$ | RMSEP (%) | RMSEC&P (%) |
| --- | --- | --- | --- |
| Baseline | 0.999 | 5.25 | 2.81 |
| PLS correction with dominant factor 1 | 0.999 | 3.28 | 1.76 |
| PLS correction with dominant factor 2 | 0.999 | 2.63 | 1.43 |
| PLS correction with dominant factor 3 | 0.999 | 2.33 | 1.27 |

Generally, the RMSEC&P in all PLS models with dominant factor are better than that in the baseline model. The RMSEC&P for the best dominant factor model with PLS correction is 1.27%, less than half of 2.81% in the baseline model, proving the proposed model with dominant factor 3 performs much better. Specifically, compared with the baseline model, the PLS models with dominant factor have a lower RMSEP value while the same high calibration quality was remained, as listed in Table 3. The lower RMSEP value indicates that the proposed dominant factor based PLS models are more robust than the baseline model for measuring samples with matrix out of the calibration set. In the PLS model with dominant factor 3, the relative errors were significantly reduced compared with the baseline model. For example, the relative error was decreased from 15.48% to 5.08% for ZBY907 and from 5.10% to 3.30% for ZBY927. The physical background of the PLS model with dominant factor plays an important role in making the model more reliable for samples with wider Cu concentration range. Compared with the dominant factor model, the PLS correction is also very effective to improve the final results, which can be confirmed by the decrease of the RMSEC&P. The PLS correction models have a much accurate result for the calibration sample set ($R^2$=0.999). However, for dominant factor 2 and 3 cases, the combination of dominant factor and PLS correction did not yield a better prediction result with a larger RMSEP value. Such results indicate that the linear correlation of PLS approach might be not able to sufficiently compensate for non-linear effect due to plasma properties fluctuations and the noise signals overfitting might make the model less accurate in predicting sample with species concentration out of the calibration set. Additionally, since empirical equation and curve fitting technology were applied in extracting the dominant factor, there could be some uncertainty in the dominant factor itself, which influence the final results. It was also found that the residual PLS correction model has the same trend as the corresponding dominant factor model for prediction. It is believed that the more accurate dominant factor model be established, the more accurate final PLS correction model results be obtained. With the development of laser induced plasma understanding, the proposed residual correction PLS model will be further improved easily. Dominant factor 3 is regarded as the best because it modelled both non-linear self-absorption and inter-element interference effects. The final results listed in Table 3 confirm the assumption. This can be further explained as follows.

Figure 10 shows the target residual needs corrected by PLS approach for the four samples for model validation. As shown, the target residual which needs to be corrected by PLS method decreases from the dominant factor 1 to dominant factor 3 model and is much smaller than that of the baseline model (the real Cu concentration in the sample, normally larger than 55%). Basically, PLS method applies linear correlation between the full spectral data and the residual to fit the non-linear relation between them. Therefore, as more non-linear effects are modelled explicitly, leaving less residual, which is non-linearly related to the spectra, to be corrected by PLS approach, this should offer better results for the combination model. In addition, the residuals for the calibration sample set share the same trend as the validation set, although they were not listed.

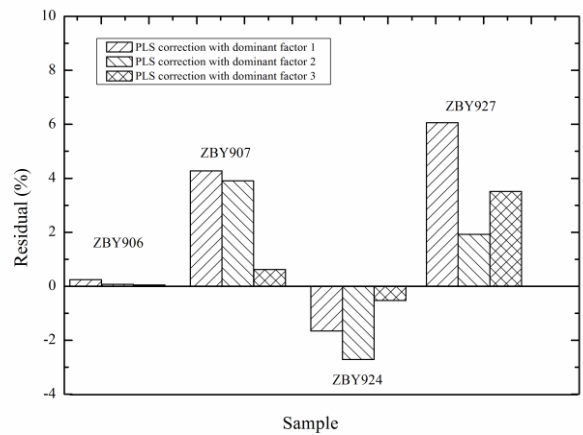

**Figure 10. Target values of different PLS models**

In essence, compared with the conventional PLS, the proposed PLS method purposely enlarges the weight of the characteristic line (570.024 nm) by explicitly extracting the dominant factor, which, in turn, reduced the possibility of noise overfitting problem and yielded better model results. This can be further demonstrated as follows.

Regression coefficient is used to represent the importance of an X-variable for Y in PLS modelling [33]. Larger PLS-regression coefficient of $x_j$ means that $x_j$ is more significant for the modeling of Y. Therefore, the intensity point with larger regression coefficient plays a more important role in determining the elemental concentration.

Figure 11 shows regression coefficients for three main points (569.943 nm, 570.069 nm, 570.195 nm) consisting the profile of the characteristic line of Cu(I) at 570.024 nm. As seen, the absolute values of regression coefficients of the PLS correction model are much smaller than those in baseline PLS model, which indicates that these spectral points are less important in explaining the concentration. This is because the concentration information contained in line intensity at 570.024 nm has been directly extracted in dominant factor. Ideally, the regression coefficients should be very close to zero for the residual correction PLS models. That is, the weight of the dominant factor is actually much larger than that of PLS calculation values in the proposed model. This helps to reduce the influence of overfitting in the final results because the dominant factor explains the major part of concentration. Besides, the regression coefficients become negative after the dominant factor extraction, which might result from the fact that the present dominant factors overly extracted concentration information from the characteristic line at 570.024 nm.

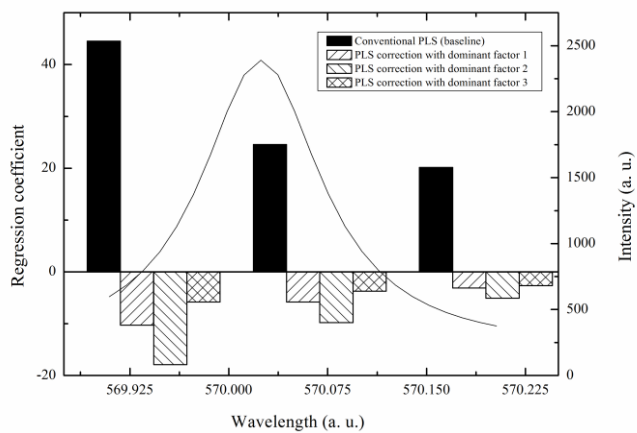

**Figure 11. Regression coefficients for the three points consisting the profile of Cu(I) line at 570.024 nm**

## 5 Conclusions

According to the deviation evolution of the characteristic intensity in LIBS measurement, the intensity of the characteristic lines may not be accurate enough to reflect the measured element concentration, while it still contains the most-related information for element concentration measurement. Linear relation between the characteristic line intensity, non-linear self-absorption effect, and inter-element interference were taken to extract the dominant factor to determining the major concentration. Based on the extracted dominant factor, a residual correction PLS model was proposed to further minimize the model prediction errors. .

Three dominant factors were extracted and the simplest one applies the linear relation between the measured element concentration and its characteristic line intensity. Self-absorption effect was then explicitly modelled to establish the dominant factor 2 model. Based on the self-absorption model, the correlation between the residuals and other elemental emission lines were used to model the inter-element interference effect for dominant factor 3. The residuals of the dominant factors were further corrected by PLS using the abundant spectral information to compensate for the plasma parameter fluctuations. As the non-linear self-absorption and inter-element interference were modelled properly, the results show that the prediction accuracy was improved greatly. Compared with the baseline model, RMSEP decreased from 5.25% to 2.33% in the dominant factor 3 model combined with PLS correction while $R^2$ remained as high as 0.999. The RMSEC&P in the dominant factor 3 model with PLS correction was 1.27%, which is much smaller than the baseline model and indicates a great improvement. The overall performance of the proposed model is the best among the models discussed in the present paper. Since the dominant factor was extracted based on the physical principle of plasma, with further improvement on plasma physics understanding, the new PLS model can be easily improved with better prediction accuracy.

The essence of the dominant-factor models was further demonstrated. Since less residual errors instead of the full concentration values were required by the linear correlation PLS approach and non-linear effects have been explicitly taken in the dominant factor, the proposed model is able to predict the sample element concentration more accurately by enlarging the weight of characteristic line intensity and reducing the noise overfitting errors.

## Acknowledgement

The authors acknowledge the financial support from the governmental "863" project (NO. 2006BAA03B04) and "973" project (NO. 2010CB227006).